\documentclass{PoS}

\usepackage{hep}
\usepackage{epsfig,multicol}
\usepackage{subfigure}
\usepackage{wrapfig}

\newcommand{\bs}[1]{\boldsymbol{#1}}
\def\d{\mathrm{d}}

\title{Probing the theoretical description of central exclusive production}

\ShortTitle{Probing the theoretical description of central exclusive production}

\author{\speaker{Tim Coughlin}%
         \thanks{This work was supported by the UK Science and Technology Facilities Council (STFC).}\\
        University College London\\
        E-mail: \email{coughlin@hep.ucl.ac.uk}}

\author{Jeffrey Forshaw\\
        University of Manchester, UK  \\
        E-mail: \email{jeff.forshaw@manchester.ac.uk}}

\abstract{We investigate the theoretical description of the central
  exclusive production process, $h_1 + h_2 \to h_1+X+h_2$. Taking
  Higgs production as an example, we compute the subset of next-to-leading order corrections sensitive to the Sudakov factor appearing in the process. Our results agree with those originally
  presented by Khoze, Martin and Ryskin except that
the scale appearing in the Sudakov factor, $\mu=0.62 \sqrt{\hat{s}}$,
should be replaced with $\mu=\sqrt{\hat{s}}$, where $\sqrt{\hat{s}}$
is the invariant mass of the centrally produced system. We show that the
replacement leads to approximately a factor 2 suppression in the
cross-section for central system masses in the range
100--500~GeV.}

\FullConference{XVIII International Workshop on Deep-Inelastic Scattering and
Related Subjects\\
                 April 19 -23, 2010\\
                 Convitto della Calza, Firenze, Italy}

\begin{document}

\section{Introduction}

At hadron colliders, in events producing high transverse momentum particles in the central rapidity region, the colliding particles usually break up. However, in a small fraction of such events the colliding hadrons remain intact and scatter through small angles. This type of production is known as central exclusive production (CEP):
\begin{align}
	h_1(p_1) +h_2(p_2) &\to h_1(p_1') \oplus X \oplus h_2(p_2')~,
\end{align}	
where the $\oplus$ denote rapidity gaps between the outgoing hadrons and the central system $X$ (see~\cite{Albrow:2010yb} for a review).


If the outgoing hadron momenta are measured, by adding detectors far down the beam-pipe, it is possible to reconstruct the four-momentum of the central system $X$. In addition, the process possesses a $J^{PC}=0^{++}$ selection rule~\cite{Khoze:2000jm}. Thus CEP offers a method to measure both the mass of $X$~\cite{Albrow:2000na} (with a resolution of $\sim 2\textrm{~GeV}$ per event~\cite{Albrow:2008pn}) and its spin-parity properties~\cite{Kaidalov:2003fw}. 

Photon pairs~\cite{:2007na}, di-jets~\cite{Aaltonen:2007hs} and $\chi_c$ particles~\cite{Aaltonen:2009kg} produced via the CEP mechanism have now been observed at the Tevatron and there are groups within both the ATLAS and CMS collaborations actively seeking to observe these events at the LHC~\cite{Albrow:2008pn}.

The CEP process has previously been calculated in perturbative QCD by the Durham group~\cite{Khoze:2000jm,Khoze:2001xm,Khoze:2000cy,Kaidalov:2003ys}. We perform an independent calculation of the next-to-leading order corrections, in the case of Higgs production, which are sensitive the Sudakov factor appearing in their result. Our finding is that the Durham group's result must be modified and we access the impact of this modification on predictions for the LHC and Tevatron~\cite{Coughlin:2009tr}.


\section{The Durham model} \label{sec:Durham}



\begin{figure}[bh]
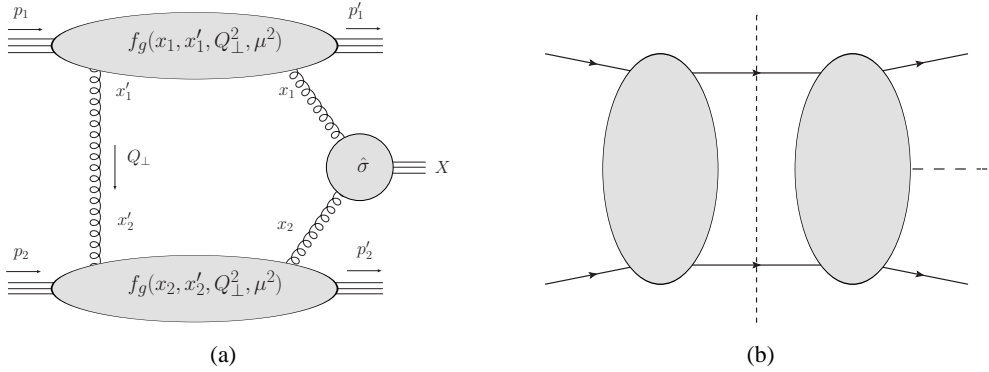

	\center
	\subfigure[]{\label{fig:CEPschematic}\includegraphics[width=0.39\textwidth]{FeynmanDiagrams/CEPschematic/CEPschematic.epsi}} \hspace{0.07\textwidth}
	\subfigure[]{\label{fig:CEPcuts}\includegraphics[width=0.39\textwidth]{ExtraDiagrams/cuts/1.epsi}} 
	\caption{(a) Schematic form of the CEP amplitude. (b) Cut of the $qQ\to q \oplus H \oplus Q$ amplitude sensitive to the Sudakov factor.}
\end{figure}


The calculation of the CEP process by the Durham group is represented schematically in figure~\ref{fig:CEPschematic}. The protons exchange a two gluon system, which must be in a colour singlet state in order that the protons remain intact. Two of the gluons then fuse to produce the central system, $X$. The cross-section is assumed to factorise in the following way~\cite{Khoze:2000cy,Khoze:2001xm}:
\begin{align}
	\frac{\partial \sigma}{\partial \hat{s} \, \partial y \, \partial \boldsymbol{p}_{1\perp}'^2 \partial \boldsymbol{p}_{2\perp}'^2} &= S^2 e^{-b(\boldsymbol{p}_{1\perp}'^2+\boldsymbol{p}_{2\perp}'^2)} \frac{\partial \mathcal{L}}{\partial \hat{s} \, \partial y} \d \hat{\sigma}(gg\to X) \;.\label{eq:Durhamdsig}
\end{align}
Where $\hat{s}$ and $y$ denote the central system invariant mass and rapidity respectively. The important piece of this expression, for our analysis, is the effective luminosity, which is given by
\begin{align}
	\frac{\partial \mathcal{L}}{\partial \hat{s}\, \partial y} &= \frac{1}{\hat{s}} \left( \frac{\pi}{N^2-1}\int \! \frac{\d\boldsymbol{Q}_\perp^2}{\boldsymbol{Q}_\perp^4} f_g(x_1,x_1',\boldsymbol{Q}_\perp^2,\mu^2) f_g(x_2,x_2',\boldsymbol{Q}_\perp^2,\mu^2)  \right)^2 \;. \label{eq:EffectiveLumi}
\end{align}
The $f_g$ are skewed, unintegrated, gluon distribution functions. Due to the kinematics of the process the amplitude is dominated by the region $x_i' \ll x_i$ and in this regime these distributions may be related to the conventional, integrated, gluon density~\cite{Martin:2001ms,Khoze:2000cy}:
\begin{align}
	f_g(x,x',\boldsymbol{Q}_\perp^2,\mu^2) \approx R_g \frac{\partial}{\partial \ln \boldsymbol{Q}_\perp^2} \left(  \sqrt{T(\boldsymbol{Q}_\perp,\mu)} x g(x,\boldsymbol{Q}_\perp^2) \right) \;,
\end{align}
with $R_g \approx 1.2(1.4)$ at the LHC(Tevatron)\footnote{For a LHC running at 14~TeV.}~\cite{Shuvaev:1999ce,Khoze:2001xm}. The $f_g$ distributions also include a Sudakov factor, which resums logarithmically enhanced soft and collinear virtual corrections and accounts for the fact that real radiation from the process is forbidden~\cite{Martin:2001ms}:
\begin{align}
	T(\boldsymbol{Q}_\perp,\mu) &= \textrm{exp}\left(  -\int_{\boldsymbol{Q}_\perp^2}^{\hat{s}/4} \! \frac{\d k_\perp^2}{k_\perp^2}\frac{\alpha_s(k_\perp^2)}{2\pi} \int_0^{1-\Delta} \! \d z \; \left[  z P_{gg}(z) +\sum_q P_{qg}(z)  \right]  \right) \; . \label{eq:DurhamSudakov}
\end{align}
For the parameters entering this expression the Durham group find~\cite{Kaidalov:2003ys}:
\begin{align}
	\Delta &= \frac{k_\perp}{k_\perp +\mu}  \;, \qquad
	\mu = 0.62 \sqrt{\hat{s}} \;. \label{eq:DurhamMU} 
\end{align}
We find that this result is incorrect and that instead one should set $\mu= \sqrt{\hat{s}}$. In the next section we describe the calculation which leads us to this conclusion.

\section{Next-to-leading order calculation and cross-section predictions}\label{sec:FullNLO}

Because we are dealing with colour singlet exchange in the high energy limit, the imaginary part of the amplitude dominates. We may therefore calculate the next-to-leading order corrections by application of the Cutkosky rules. Furthermore, since we are only interested in probing the Sudakov factor of the Durham resullt, we may limit ourselves to the quark-quark scattering channel and the cut diagram shown in figure~\ref{fig:CEPcuts} (and the diagram in which the Higgs attaches to the left of the cut\footnote{A full list of the diagrams calculated and the argument for neglecting other cuts may be found in~\cite{Coughlin:2009tr}.}).



After performing the loop integrals and subtracting the infrared divergent terms associated with the perturbative expansion of the parton distribution functions, we find for the next-to-leading order contribution to the amplitude~\cite{Coughlin:2009tr}
\begin{align}
	A_{\textrm{NLO}} &\approx \int \! \frac{d\bs{Q}_\perp^2}{\bs{Q}_\perp^4} A_0(m_H) \ln(T(\bs{Q}_\perp,m_H)) \; , \label{eq:MyResult}
\end{align}
where we have neglected terms not enhanced by at least one large logarithm and those suppressed by a power of $\bs{Q}_\perp$. The constant $A_0$ is related to the lowest order amplitude as $A_{\textrm{LO}} \approx \int \! d\bs{Q}_\perp^2 A_0(m_H) / \bs{Q}_\perp^4$,
where again terms suppressed by $\bs{Q}_\perp$ are neglected. Comparing equation~(\ref{eq:MyResult}) with the Durham result (equations~(\ref{eq:EffectiveLumi})-(\ref{eq:DurhamMU})), we see that the general form is correct, but one must replace $\mu=0.62 \sqrt{\hat{s}}$ with $\mu= \sqrt{\hat{s}}$.

\begin{figure}[tb]
	\centering
	\includegraphics[width=0.55\textwidth]{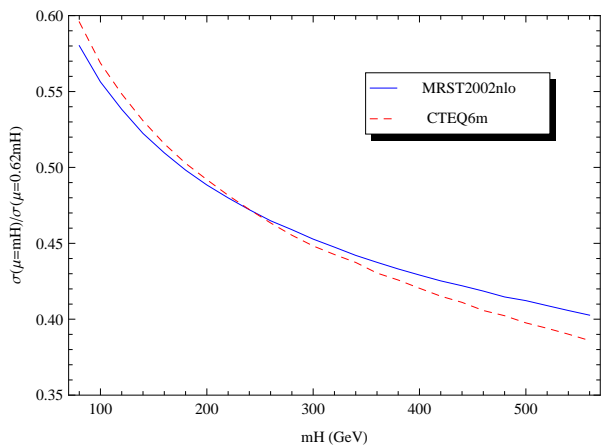}
	\caption{Ratio of the cross-section for central exclusive Higgs production at the LHC evaluated with the scale in the Sudakov factor set to $\mu=m_H$ divided by the cross-section with the scale set to $\mu=0.62 m_H$, plotted as a function of the Higgs mass. The solid blue and dashed red lines were generated using MRST2002nlo~\cite{MRST} and CTEQ6m~\cite{CTEQ} parton distributions respectively.} \label{fig:AndysNumbers}
\end{figure}

We may now assess the impact that our modification of the Sudakov factor has on predictions of the central exclusive cross-section. Taking, as an example, the cross-section for central exclusive Higgs production at the LHC, with 14~TeV centre-of-mass energy we compute the cross-section, using the ExHuME Monte Carlo generator~\cite{Monk:2005ji}, placing no cuts on the final-state particles. The results are shown in figure~\ref{fig:AndysNumbers}, for two different parton distribution functions \cite{MRST,CTEQ}. We observe a suppression of the cross-section, relative to the predictions of the Durham group, by a factor $\sim 2$ which increases with increasing Higgs mass. 

\section{Conclusions}\label{sec:conclusions}

We have studied the cross-section for central exclusive Higgs boson production, using QCD perturbation theory. We largely confirm the calculation previously performed by the Durham group, except that we disagree as to the precise form of the Sudakov factor which enters. Using the Sudakov factor that we propose leads to a suppression of the central exclusive production cross-section at the LHC by approximately a factor of two relative to the earlier predictions, for Higgs boson masses in the range 100--500~GeV.

As a point of further study, it would be interesting to assess the impact on predictions for other processes and in particular on the central exclusive production of dijets at the Tevatron, for which data exist. We do not expect to find any disagreement with the data. In fact, the reduced suppression at lower masses suggests that agreement with the data may even be slightly improved. However, one must always remember that the theoretical uncertainty on other parts of the calculation, for example the gap survival factor and unintegrated pdfs, is expected to be comparable in size to the effect induced by the change in the Sudakov factor that we have been focussing on.  

We note that the fixed-order corrections we have computed form a subset of the full next-to-leading order corrections to central exclusive Higgs production, offering the possibility of extending the theoretical description of the process to this order.


  


  

\bibliographystyle{JHEP}
\bibliography{DIS2010-Tim_Coughlin}

\end{document}